\documentclass[12pt]{iopart}

\usepackage{graphicx}
\usepackage{iopams}
\usepackage{cite}

\newcommand{\thetaf}[4]{%
\theta\!\left[ #1 \atop #2 \right]\!\left(#3,#4\right)}

\begin{document}

\title{Lattice Laughlin states on the torus from conformal field theory}
\author{Abhinav Deshpande$^{1,2}$ and Anne E B Nielsen$^{1,3}$}
\address{$^1$ Max-Planck-Institut f{\"u}r Quantenoptik,
Hans-Kopfermann-Stra{\ss}e 1, D-85748 Garching, Germany}
\address{$^2$ Department of Physics, Indian Institute of Technology Kanpur, Kanpur, UP- 208016, India}
\address{$^3$ Department of Physics and Astronomy, Aarhus University, Ny Munkegade 120, DK-8000 Aarhus C, Denmark}

\begin{abstract}
Conformal field theory has turned out to be a powerful tool to derive two-dimensional lattice models displaying fractional quantum Hall physics. So far most of the work has been for lattices with open boundary conditions in at least one of the two directions, but it is desirable to also be able to handle the case of periodic boundary conditions. Here, we take steps in this direction by deriving analytical expressions for a family of conformal field theory states on the torus that is closely related to the family of bosonic and fermionic Laughlin states. We compute how the states transform when a particle is moved around the torus and when the states are translated or rotated, and we provide numerical evidence in particular cases that the states become orthonormal up to a common factor for large lattices. We use these results to find the $S$-matrix of the states, which turns out to be the same as for the continuum Laughlin states. Finally, we show that when the states are defined on a square lattice with suitable lattice spacing they practically coincide with the Laughlin states restricted to a lattice.
\end{abstract}

\begin{indented}\item[]
\end{indented}

\section{Introduction}

In certain two-dimensional systems consisting of many interacting bosons or fermions, it is possible to have quasi-particles with unusual properties like fractional charge and an exchange statistics that is neither bosonic nor fermionic. The properties of such topological systems are very different from other types of matter, and they are therefore attracting a lot of attention.

The quantum states appearing in connection with the fractional quantum Hall (FQH) effect constitute important examples of topological states, because they can be realized experimentally in semiconductor devices under demanding conditions, and because many of the states are believed to be described by simple analytical wave functions \cite{wf,hierarchy1,hierarchy2,haldane85a,haldane85b,cf,CFTstates,chung07,hermanns,hermanns2}. The interest in topological systems has triggered the question whether FQH physics can occur in other settings, and in the last years there have been several proposals for how to obtain FQH behaviour in lattice systems. This research allows for the investigation of new aspects of FQH physics and paves the way towards new possibilities for realizing the effect experimentally \cite{jaksch2003,PRA.70.041603,PRL.94.086803,hafezi2008,PhysRevLett.96.180407,PhysRevA.82.043629,cooper,yao,nc,sterdyniak}.

There are mainly two strategies to obtain lattice models displaying FQH physics. One of them is to engineer a band structure that is reminiscent of a Landau level, ensure that the band is partially filled, and add interactions between the particles \cite{neupert,sheng,wang,regnault}. The other strategy is to start from an analytical FQH wave function in the continuum, modify it to a lattice wave function, and then analyze its properties and use analytical tools to derive a Hamiltonian for which the state is the ground state \cite{schroeter,thomale,kapit,prl,nc,son,greiter,genq,sun,bondesan,nielsen,glasser15,rodriguez}. Both approaches have been shown numerically to be successful.

It has long been known that there is a connection between FQH physics and conformal field theory (CFT) and that a number of continuum FQH states can be expressed as conformal blocks of certain CFT correlators \cite{CFTstates}. This observation is also helpful for finding analytical expressions of FQH states with periodic boundary conditions \cite{hermanns}. More recently, the CFT formulation has turned out to be very fruitful for constructing FQH models in lattice systems with analytical wave functions and corresponding parent Hamiltonians \cite{idmps,su2k,prl,nc,son,genq,sun,bondesan,nielsen,glasser15,rodriguez}.

A natural way to transform a wave function in the continuum into a lattice wave function is to keep the analytical expression for the state, but to let the state be a sum over all possible distributions of the particles on a lattice rather than an integral over all possible positions of the particles in the plane. Part of the trick used to derive the parent Hamiltonians in \cite{su2k,prl,son,genq,sun,bondesan,nielsen,glasser15}, however, is not just to do this, but to make an additional change. The Gaussian factor appearing in the continuum FQH wave functions is obtained by including a uniform background charge in the CFT correlator, and the additional change is to restrict the background charge to the same lattice as the particles. This has the further advantage that the CFT correlator takes a particularly simple form: for a lattice with $N$ sites, it is the expectation value of a product on $N$ CFT operators without any other factors. The analytical expression for the lattice states then differs slightly from the analytical expression for the continuum states, but at least as long as the distribution of the lattice sites is not too far from uniform, this does not change the physical properties of the states significantly.

Apart from a study of the chiral spin liquid Kalmeyer-Laughlin state in \cite{torus}, the investigations of lattice FQH behaviour obtained from CFT as described above has so far been concerned with systems having periodic boundary conditions in at most one direction, because the case of periodic boundary conditions in both directions is more complicated to handle in CFT. It is, however, very desirable to be able to investigate two-dimensional models with periodic boundary conditions in both directions, since the topology of the space on which a state is defined plays an important role for topological systems. In particular, several computational methods have been developed in recent years to determine particular topological properties of quantum systems \cite{kitaev,levin,jiang,quasiparticle,cincio,zaletel,tu13,orus}, but many of these rely on the ability to define the investigated systems on surfaces with periodic, or more generally twisted, boundary conditions. Systems defined on a torus are also appealing because the lack of a boundary makes it easier to reliably investigate the bulk properties of the states.

In the present paper, we take steps towards generalizing the previous work on lattice FQH physics obtained using CFT techniques to the torus. Our starting point is a CFT correlator that has previously \cite{genq} been used to derive lattice models with Laughlin-like ground states on the plane and cylinder. By evaluating the same correlator on the torus we obtain analytical expressions for a family of CFT states that is closely related to the family of bosonic and fermionic Laughlin states on the torus. We investigate the boundary conditions of these states when a single particle is moved around the torus, and we study how the states on particular lattices transform under collective rotation and translation operations. We provide numerical evidence in particular cases that the states become orthonormal up to a common factor for large lattices, and from this and the properties under rotation we conclude that the $S$-matrix of the states is the one expected for the Laughlin states. The $ij$'th entry of the $S$-matrix gives the phase factor acquired by the wave function when an anyon of type $i$ is moved adiabatically in a closed path around an anyon of type $j$. The result hence shows that the braiding properties of the anyons are those of the Laughlin anyons. Finally, we show that, for the case of a square lattice, the analytical expressions of the states practically coincide with the analytical expressions of the corresponding Laughlin states on the torus.

\section{Lattice Laughlin states on the plane from CFT}

We first briefly recall the construction of lattice Laughlin states on the plane from CFT proposed in \cite{genq}. The starting point is the CFT correlator
\begin{eqnarray}\label{CFTcorplane}
\langle \phi_{n_1}(z_1,\bar{z}_1) \phi_{n_2}(z_2,\bar{z}_2) \cdots \phi_{n_N}(z_N,\bar{z}_N) \rangle_\mathrm{P},
\end{eqnarray}
where $\langle\ldots\rangle$ denotes the vacuum expectation value, the subscript P stands for plane, and
\begin{eqnarray}
\phi_{n_j}(z_j,\bar{z}_j)={:\rme^{\rmi \frac{qn_j -1 }{\sqrt{q}} \varphi(z_j,\bar{z}_j)}:}
\end{eqnarray}
is a vertex operator \cite{CFTbook}. Here, $:\ldots:$ means normal ordering, $q$ is a positive integer, $n_j\in\{0,1\}$, $z_j$ is a complex number, $\bar{z}_j$ is the complex conjugate of $z_j$, and $\varphi(z_j,\bar{z}_j)$ is the field of a free, massless boson compactified on a circle of radius $R=\sqrt{q}$.

CFT correlators can be decomposed \cite{CFTbook,block} into conformal blocks that are functions of the holomorphic coordinates $z_j$, but not of the antiholomorphic coordinates $\bar{z}_j$. This decomposition allows us to write \eref{CFTcorplane} in the form
\begin{eqnarray}\label{cbplane}
\langle \phi_{n_1}(z_1,\bar{z}_1) \phi_{n_2}(z_2,\bar{z}_2) \cdots \phi_{n_N}(z_N,\bar{z}_N) \rangle_\mathrm{P} = \sum_k c_k^\mathrm{P} f_k^\mathrm{P}(n_j,z_j) \overline{f_k^\mathrm{P}(n_j,z_j)},
\end{eqnarray}
where $f_k^\mathrm{P}(n_j,z_j)$ are the conformal blocks and $c_k^\mathrm{P}$ are numbers that do not depend on $z_j$ or $\bar{z}_j$.

On the plane, there is only one conformal block $f_1^\mathrm{P}(n_j,z_j)$, and we use this conformal block to define a wave function on a two-dimensional lattice as follows. We interpret $(\mathrm{Re}(z_j),\mathrm{Im}(z_j))$ as the position in the plane of the $j$'th lattice site and $n_j$ as the number of particles on the $j$'th lattice site. We then write
\begin{eqnarray}\label{Laughlin_plane}
|\psi_\mathrm{P}\rangle\equiv\sum_{n_1,n_2,\ldots,n_N} \psi_\mathrm{P}(n_1,n_2,\ldots,n_N) |n_1,n_2,\ldots,n_N\rangle,
\end{eqnarray}
where
\begin{eqnarray}\label{eqgenqplane}
\psi_\mathrm{P}(n_1,n_2,\ldots,n_N) \propto f_1^\mathrm{P}(n_j,z_j) = \delta_n \chi_n \prod_{i<j} (z_i-z_j)^{qn_in_j-n_i-n_j+1/q}.
\end{eqnarray}
The factor $\delta_n$ is defined as
\begin{eqnarray}
\delta_n=\left\{\begin{array}{cl}
1 & \textrm{for } \sum_in_i=N/q, \\
0 & \textrm{otherwise},
\end{array}\right.
\end{eqnarray}
and it fixes the lattice filling factor $\sum_in_i/N$ to $1/q$. Since \eref{cbplane} does not fix the phase of the conformal block, we include an unspecified phase factor $\chi_n$. Note that $\chi_n$ cannot depend on the lattice positions due to the requirement that the conformal block is holomorphic, but it may depend on $n_j$.

As shown in \cite{genq}, the factor $\prod_{i<j} (z_i - z_j)^{-n_i-n_j}$ approaches the Gaussian factor of the Laughlin state with filling factor $1/q$ up to a phase factor when $N\to\infty$ if the lattice is regular (with the area per lattice site being the same for all lattice sites) and the boundary of the lattice is a circle. The phase factor can be transformed away if desired, and in practice, $N\approx100$ is already enough to obtain very good agreement \cite{genq}. We thus observe that \eref{eqgenqplane} is practically the Laughlin state with filling factor $1/q$, except that the possible positions of the particles are restricted to a set of lattice sites. In \cite{genq}, it has also been shown that these states have the same topological entanglement entropy $-\ln(q)/2$ as the continuous Laughlin states.

\section{Conformal blocks on the torus}

In this section, we derive the conformal blocks of the correlator \eref{CFTcorplane} on the torus (the final result can be found in section \ref{sec:cb}). We first define the torus and the coordinates, we are using. Let $\omega_1$ and $\omega_2$ be two complex numbers. The modular parameter is defined as $\tau=\omega_2/\omega_1$, and we assume that $\mathrm{Im}(\tau)>0$. The torus is then the parallelogram spanned by $\omega_1$ and $\omega_2$ with periodic boundary conditions. In other words, if $n$ and $m$ are integers, and $a$ and $b$ are real numbers in the interval $[0,1[$, then we identify the point $(n+a)\omega_1+(m+b)\omega_2$ with the point $a\omega_1+b\omega_2$. Let $z$ be a point within the parallelogram. We then define the scaled coordinate $\xi=z/\omega_1$, which we shall use throughout the paper.

We would like to determine
\begin{eqnarray}\label{define_state}
\langle \phi_{n_1}(\xi_1,\bar{\xi}_1) \phi_{n_2}(\xi_2,\bar{\xi}_2) \cdots \phi_{n_N}(\xi_N,\bar{\xi}_N) \rangle_\mathrm{T} = \sum_k c_k f_k(n_j,\xi_j) \overline{f_k(n_j,\xi_j)},
\end{eqnarray}
where $\mathrm{T}$ stands for torus. Our starting point for evaluating \eref{define_state} is the following relation for the correlator of a product of vertex operators on the torus \cite{cortorus1} (see also \cite{cortorus2})
\begin{eqnarray}\label{correlatortor}
\fl\left \langle :\rme^{\rmi (\nu_1 \varphi(\xi_1) + \bar{\nu}_1 \bar{\varphi}(\bar{\xi}_1)) } : :\rme^{\rmi (\nu_2 \varphi(\xi_2) + \bar{\nu}_2 \bar{\varphi}(\bar{\xi}_2)) } : \cdots :\rme^{\rmi (\nu_N \varphi(\xi_N) + \bar{\nu}_N \bar{\varphi}(\bar{\xi}_N)) } :\right \rangle_\mathrm{T} \nonumber\\ = \frac{\delta_\nu \delta_{\bar{\nu}} }{|\eta(\tau)|^2 } \sum_{(p, \bar{p})\in \Gamma } A_0^p (\xi_i, \nu_i) \overline{A_0^p (\xi_i, \nu_i)}.
\end{eqnarray}
On the left hand side of \eref{correlatortor}, $\varphi(\xi_j)$ ($\bar{\varphi}(\bar{\xi}_j)$) is the holomorphic (antiholomorphic) part of the field $\varphi(\xi_j,\bar{\xi}_j)=\varphi(\xi_j)+\bar{\varphi}(\bar{\xi}_j)$ of the free, massless boson and the numbers ($\nu_j$, $\bar{\nu_j}$) $\in \Gamma$, where $\Gamma$ is the lattice of the momenta consisting of the elements $(p,\bar{p})$ with
\begin{eqnarray}
p = \frac{n}{R} + \frac{m R}{2}, \quad \bar{p} = \frac{n}{R} - \frac{m R}{2}, \quad n,m \in \mathbb{Z}.
\end{eqnarray}
Here, $R=\sqrt{q}$ is the compactification radius mentioned above. For the correlator \eref{define_state}, the relevant values of $(\nu_j,\bar{\nu}_j)$ are $(-1/\sqrt{q},-1/\sqrt{q})$ and $((q-1)/\sqrt{q},(q-1)/\sqrt{q})$, which are obtained for $m=0$ and $n=-1$ or $n=q-1$, respectively.

On the right hand side of \eref{correlatortor}, $\delta_\nu=1$ if $\sum_j\nu_j=0$ and $\delta_\nu=0$ otherwise,
\begin{eqnarray}\label{dedekind}
\eta(\tau)\equiv\rme^{\rmi \pi \tau / 12 } \prod_{n=1}^{\infty}(1-\rme^{2 \pi \rmi \tau n})
\end{eqnarray}
is the Dedekind eta function,
\begin{eqnarray}\label{Adef}
A_0^p (\xi_i, \nu_i) = \rme^{\rmi \pi \tau p^2 + 2\pi \rmi p \sum_i \xi_i \nu_i } \prod_{i < j} E (\xi_i - \xi_j , \tau)^{\nu_i \nu_j},
\end{eqnarray}
and $\overline{A_0^p (\xi_i, \nu_i)}$ is the function obtained by taking the complex conjugate of $A_0^p$ and replacing $\nu_i$ with $\bar{\nu}_i$ and $p$ with $\bar{p}$. $E$ is defined as
\begin{eqnarray}
E(\xi_i - \xi_j, \tau) = \frac{\theta_1 (\xi_i - \xi_j, \tau) }{\partial_{\xi}{\theta_1 (\xi, \tau)} |_{\xi = 0} }, \qquad \theta_1 (\xi, \tau) \equiv \thetaf{1/2}{1/2}{\xi}{\tau},
\end{eqnarray}
where the Riemann theta function is given by
\begin{eqnarray}\label{thetafunction}
\thetaf{a}{b}{\xi}{\tau}=\sum_{n \in \mathbb{Z} } \rme^{\rmi\pi \tau(n+a)^2 + 2\pi \rmi (n+a)(\xi + b)}.
\end{eqnarray}

The only part of \eref{correlatortor} that does not immediately factorize into holomorphic and corresponding antiholomorphic parts is the factor
\begin{eqnarray}\label{expsimplify}
\fl\sum_{(p, \bar{p} ) \in \Gamma } \exp{\left(
\rmi\pi\tau p^2+2\pi\rmi p \sum_i\xi_i\nu_i
-\rmi\pi\bar{\tau}\bar{p}^2-2\pi\rmi\bar{p}\sum_i\bar{\xi}_i\bar{\nu}_i \right)}\nonumber\\
= \sum_{n,m \in \mathbb{Z} } \exp{ \left[i \pi \tau q \left(n/q + m/2\right)^2 + 2\pi i \left(n/q + m/2 \right) \xi/\sqrt{q} \right] } \nonumber \\
   \times\exp{\left[ -i \pi  \bar{\tau} q\left(n/q - m/2\right)^2 - 2\pi i \left(n/q - m/2 \right) \bar{\xi}/\sqrt{q} \right]},
\end{eqnarray}
where we have defined $\xi\equiv q\sum_i\xi_i\nu_i$. We shall consider the cases $q$ even and $q$ odd separately in the following.

\subsection{Even $q$}
When $q$ is even, $(n+mq/2)$ and $(n-mq/2)$ are integers and differ by a multiple of $q$. Therefore, upon division by $q$, they give the same remainder $l$. We can thus write
\begin{eqnarray}\label{ldef}
p = \sqrt{q}(r + l/q), \quad \bar{p} = \sqrt{q}(s + l/q), \nonumber \\
r = \left\lfloor \frac{n + mq/2}{q} \right\rfloor, \quad s = \left\lfloor \frac{n - mq/2}{q} \right\rfloor, \\
l = (n + mq/2, \textrm{mod} \ q) = (n - mq/2, \textrm{mod} \ q), \nonumber
\end{eqnarray}
where $l \in \{0,1,..q-1\}$ and $r,s \in \mathbb{Z}$ without any restriction. That is, for a fixed $r$, the summation over $s$ is over the whole set $\mathbb{Z}$. Therefore, \eref{expsimplify} becomes
\begin{eqnarray}
\sum_{l \in \{0,1.. q-1 \} } \sum_{r \in \mathbb{Z} } \rme^{\rmi \pi q \tau \left(r + \frac{l}{q} \right)^2 + 2\pi \rmi \left(r + \frac{l}{q} \right) \frac{\xi}{\sqrt{q}} } \sum_{s \in \mathbb{Z} }
\rme^{ -\rmi \pi q \bar{\tau} \left(s + \frac{l}{q} \right)^2 - 2\pi \rmi \left(s + \frac{l}{q} \right) \frac{\bar{\xi}}{\sqrt{q}}},
\end{eqnarray}
which is a sum of factorized terms.

\subsection{Odd $q$}
When $q$ is odd, $(n + mq/2)$ and $(n-mq/2)$ can take half integer values. In this case we split the sum over $m$ in \eref{expsimplify} into a sum over even and a sum over odd $m$. When $m$ is even, we again write
\begin{eqnarray}
p = \sqrt{q}(r + l/q), \quad \bar{p} = \sqrt{q}(s + l/q), \nonumber \\
r = \left\lfloor \frac{n + mq/2}{q} \right\rfloor, \quad
s = \left\lfloor \frac{n - mq/2}{q} \right\rfloor, \\
l = (n + mq/2, \textrm{mod} \ q) = (n - mq/2, \textrm{mod} \ q) \nonumber,
\end{eqnarray}
but here we note that $r$ and $s$ are not unrestricted. Particularly, since $m$ is even, we must have
\begin{eqnarray}
\frac{p - \bar{p}}{\sqrt{q}} = m = r - s = \textrm{even}.
\end{eqnarray}
Therefore, for a fixed $r$, say even (odd), the summation over $s$ is over even (odd) integers. We can hence rewrite the sum in \eref{expsimplify} over even $m$ into
\begin{eqnarray}\label{eqset1}
\fl\sum_{l \in \{0,1 \ldots q-1 \} } \sum_{r,s \in \mathbb{Z} } \rme^{\rmi \pi q \tau \left(r + \frac{l}{q} \right)^2 -\rmi \pi q \bar{\tau} \left(s + \frac{l}{q} \right)^2 + 2\pi \rmi \left(r + \frac{l}{q} \right) \frac{\xi}{\sqrt{q}} - 2\pi \rmi \left(s + \frac{l}{q} \right) \frac{\bar{\xi}}{\sqrt{q}}} \left( \frac{1 + \rme^{\rmi\pi(r-s)}}{2} \right),
\end{eqnarray}
where the last factor is $0$ when $r$ and $s$ are of opposite parity and $1$ otherwise.

When $m$ is odd, we write instead
\begin{eqnarray}
p = \sqrt{q}(r + l/q +1/2), \quad \bar{p} = \sqrt{q}(s + l/q + 1/2), \nonumber \\
r = \left\lfloor \frac{n + (m-1)q/2}{q} \right\rfloor, \quad  s = \left\lfloor \frac{n - (m-1)q/2}{q} \right\rfloor, \\
l = (n + (m-1)q/2, \textrm{mod} \ q) = (n - (m-1)q/2, \textrm{mod} \ q). \nonumber
\end{eqnarray}
In this case, we have
\begin{eqnarray}
\frac{p - \bar{p}}{\sqrt{q}} = m = r - s = \textrm{odd},
\end{eqnarray}
and so the summation over $s$ has opposite parity to that over $r$. This allows us to rewrite the sum in \eref{expsimplify} over odd $m$ into
\begin{eqnarray}\label{eqset2}
\fl\sum_{l \in \{0,1 \ldots q-1 \} } \sum_{r,s \in \mathbb{Z} } \rme^{\rmi \pi q \tau \left(r + \frac{l}{q} + \frac{1}{2} \right)^2 -\rmi \pi q \bar{\tau} \left(s + \frac{l}{q} + \frac{1}{2} \right)^2 + 2\pi \rmi \left(r + \frac{l}{q} + \frac{1}{2} \right) \frac{\xi}{\sqrt{q}} - 2\pi \rmi \left(s + \frac{l}{q} + \frac{1}{2} \right) \frac{\bar{\xi}}{\sqrt{q}}} \left( \frac{1 - \rme^{\rmi\pi(r-s)}}{2} \right).
\end{eqnarray}
The expression in \eref{expsimplify} then equals the sum of \eref{eqset1} and \eref{eqset2}.

\subsection{Expressions for the conformal blocks}\label{sec:cb}

Collecting the above results, we conclude that the conformal blocks take the form
\begin{eqnarray}\label{cbtorus}
\fl f_{l,A}(n_j, \xi_j ) = \frac{\delta_{n}\chi_n}{\eta(\tau)} \,\thetaf{l/q+a_A}{b_A}{\sum_i \xi_i (q n_i - 1)}{q\tau}
\prod_{i < j} E(\xi_i - \xi_j, \tau)^{qn_in_j-n_i-n_j+1/q}\!,
\end{eqnarray}
where $\delta_n=1$ if $\sum_i n_i = N/q$ and $\delta_n=0$ otherwise, $\chi_n$ is an undetermined phase factor that may depend on $n_j$, but not on $\xi_j$, and we have split the index $k$ into two indices $l$ and $A$, where $l\in\{0,1,\ldots,q-1\}$. For $q$ even, $A$ can take only the value $\mathrm{I}$, and for $q$ odd,  $A\in\{\mathrm{I},\mathrm{II},\mathrm{III},\mathrm{IV}\}$. The numbers $a_A$ and $b_A$ are defined as
\begin{eqnarray}\label{aAbA}
\begin{array}{lll}
a_\mathrm{I}=0, \quad & b_\mathrm{I}=0, \quad
& (\textrm{for }q \textrm{ even or odd}),\\
a_\mathrm{II}=0, \quad & b_\mathrm{II}=1/2, \quad
& (\textrm{for }q \textrm{ odd}),\\
a_\mathrm{III}=1/2, \quad & b_\mathrm{III}=0, \quad
& (\textrm{for }q \textrm{ odd}),\\
a_\mathrm{IV}=1/2, \quad & b_\mathrm{IV}=1/2, \quad
& (\textrm{for }q \textrm{ odd}).
\end{array}
\end{eqnarray}
For $q$ even, the factors $c_{l,\mathrm{I}}$ in \eref{define_state} are all $1$, and for $q$ odd, we have $c_{l,\mathrm{I}}=c_{l,\mathrm{II}}=c_{l,\mathrm{III}}=-c_{l,\mathrm{IV}}=1/2$. There are hence $q$ conformal blocks for $q$ even and $4q$ conformal blocks for $q$ odd.

\subsection{Phase factor}\label{subsec:phase}

Let us finally discuss the relation between the phase factor $\chi_n$ for the conformal blocks on the torus and the phase factor $\chi_n$ for the conformal block on the plane. Locally the torus looks like the plane, and we can hence recover the conformal block on the plane by putting all the $N$ lattice sites close together. Note that this does not affect $\chi_n$, since $\chi_n$ does not depend on $\xi_j$. In this case,
\begin{eqnarray}
E(\xi_i - \xi_j, \tau) = \frac{\theta_1(\xi_i-\xi_j, \tau)}{\partial_\xi \theta_1(\xi, \tau)|_{\xi=0} }
\approx \xi_i - \xi_j
\end{eqnarray}
and
\begin{eqnarray}\label{thetaapp}
\thetaf{a}{b}{\sum_i \xi_i(qn_i-1)}{q\tau} \approx \thetaf{a}{b}{0}{q\tau}.
\end{eqnarray}
The right hand side of \eref{thetaapp} is zero for $a=b=1/2$, but is nonzero for the other values of $a$ and $b$ occurring in the centre-of-mass factors. In this limit, we hence have only one conformal block, satisfying
\begin{eqnarray}\label{eqphasesimple}
\fl f_k(\xi_j,n_j) \propto \delta_{n} \chi_n \prod_{i < j} (\xi_i - \xi_j)^{qn_in_j-n_i-n_j+1/q}
\propto \delta_{n} \chi_n \prod_{i < j} (z_i - z_j)^{qn_in_j-n_i-n_j+1/q}.
\end{eqnarray}
Comparing this result to \eref{eqgenqplane}, we conclude that it is natural to choose $\chi_n$ to be the same on the torus and on the plane. In the following, we shall take
\begin{eqnarray}
\chi_n=(-1)^{\sum_j(j-1)n_j}.
\end{eqnarray}
For $q=2$ this choice ensures that the state is an SU(2) singlet \cite{torus}.

\section{Wave functions from conformal blocks}\label{sec:wf}

We use the conformal blocks on the torus \eref{cbtorus} to define the wave functions
\begin{eqnarray}\label{eqstate}
\fl|\psi_{l/q+a_A,b_A}\rangle = \sum_{n_1,n_2,\ldots,n_N} \psi_{l/q+a_A,b_A}(\xi_j,n_j) |n_1,n_2, \ldots n_N\rangle,\\
\fl\psi_{l/q+a_A,b_A}(\xi_j,n_j)=
\delta_{n} \chi_n \, \thetaf{l/q+a_A}{b_A}{\sum_{i=1}^N \xi_i (qn_i - 1)}{q\tau}
\prod_{i < j} E(\xi_i - \xi_j, \tau)^{qn_in_j -n_i- n_j },\nonumber
\end{eqnarray}
where $\psi_{l/q+a_A,b_A}(\xi_j,n_j)$ is shorthand notation for $\psi_{l/q+a_A,b_A}(\xi_1,\ldots,\xi_N,n_1,\ldots,n_N)$ and we have left out some constant factors that only affect the norm of the states. Note that $|\psi_{l/q+a_A,b_A}\rangle$ is not normalized.

Since
\begin{eqnarray}
\fl\prod_{i < j}  E(\xi_i - \xi_j, \tau)^{qn_in_j -n_i- n_j }
=(-1)^{\sum_j(j-1)n_j} \prod_{i < j}  E(\xi_i - \xi_j, \tau)^{qn_in_j}  \prod_{i\neq j} E(\xi_i - \xi_j, \tau)^{-n_i},
\end{eqnarray}
we can alternatively express \eref{eqstate} as
\begin{eqnarray}\label{statep}
\fl|\psi_{l/q+a_A,b_A}\rangle=\frac{1}{M!} \sum_{\{\Xi_1,\ldots,\Xi_M\}}
\psi_{l/q+a_A,b_A}(\xi_1,\ldots,\xi_n,\Xi_1,\ldots,\Xi_M) \; a_{\Xi_1}^\dag a_{\Xi_2}^\dag \cdots a_{\Xi_M}^\dag|0\rangle,\\
\fl\psi_{l/q+a_A,b_A}(\xi_1,\ldots,\xi_N,\Xi_1,\ldots,\Xi_M)=
\thetaf{l/q+a_A}{b_A}{q\sum_{i=1}^M \Xi_i - \sum_{i=1}^N\xi_i}{q\tau} \nonumber\\
\times\prod_{i < j} E(\Xi_i - \Xi_j, \tau)^q
\prod_{i=1}^M\prod_{j=1}^N \tilde{E}(\Xi_i - \xi_j, \tau)^{-1}.\nonumber
\end{eqnarray}
Here, $M=\sum_in_i=N/q$ is the total number of particles, $a_\Xi^\dag$ is the operator that creates a particle at position $\Xi$, $|0\rangle$ is the vacuum state, and the first sum is over all possible configurations of the particles on the lattice, i.e., all $\{\Xi_1,\Xi_2,\ldots,\Xi_M\}$ for which $\Xi_j\in\{\xi_1,\xi_2,\ldots,\xi_N\}$ for all $j$ and $\Xi_i\neq \Xi_j$ for all $i$ and $j$. In addition,
\begin{eqnarray}
\tilde{E}(\Xi,\tau)\equiv\left\{\begin{array}{cl}1 & \textrm{for }\Xi=0\\
E(\Xi,\tau) & \textrm{otherwise}\end{array}\right..
\end{eqnarray}
Since $E(-\Xi,\tau)=-E(\Xi,\tau)$, we observe that the states \eref{statep} describe bosons for $q$ even and fermions for $q$ odd. Note also that if we choose two of the coordinates $\Xi_i$ and $\Xi_j$ to be the same in \eref{statep}, then the amplitude of the term is zero because $E(0,\tau)=0$. Independent of $q$, there is hence at most one particle on each site, which means that the bosons are hardcore bosons. This result reflects the property of the Laughlin states that the amplitude for two particles being at the same point is zero.

In \eref{define_state}, we chose to order the vertex operators so that the one evaluated at the point $\xi_j$ was the $j$'th vertex operator in the correlator, and in \eref{eqstate}, we chose the ordering of the $n_j$'s in the ket to follow the ordering of the vertex operators in the correlator. We could instead have chosen the ordering such that the $j$'th vertex operator was operator number $p(j)$ in the correlator, where $p$ is some bijective map from $\{1,2,\ldots,N\}$ to $\{1,2,\ldots,N\}$. In \eref{eqstate}, this would change $\chi_n$ into $(-1)^{\sum_j(p(j)-1)n_j}$, it would change the ordering in the ket, and it would change the product over $i<j$ into the product over all $i$ and $j$ for which $p(i)<p(j)$. In \eref{statep} there is, however, no longer any reference to this ordering, and $|\psi_{l/q+a_A,b_A}\rangle$ is hence invariant under reordering. This means that it does not make a difference how we choose to label the lattice sites.

\section{Boundary conditions and fluxes}
In this section, we investigate the boundary conditions of the state \eref{statep} when the $k$'th particle is moved all the way around the torus and back to its original position, i.e.\ when $\Xi_k\to \Xi_k+1$ or $\Xi_k\to \Xi_k+\tau$. We also discuss how to generalize the states to fulfil twisted boundary conditions.

\subsection{$\Xi_k\to \Xi_k+1$}

When $\Xi_k \rightarrow \Xi_k + 1$, we have $ q \sum_i  \Xi_i \rightarrow q \sum_i \Xi_i + q$. Utilizing
\begin{eqnarray}\label{thetaprop2}
\thetaf{a}{b}{\xi+n}{\tau} = \thetaf{a}{b+n}{\xi}{\tau} = \rme^{2\pi \rmi an} \thetaf{a}{b}{\xi}{\tau} \textrm{ for } n \in \mathbb{Z},
\end{eqnarray}
we conclude that the centre-of-mass factor transforms as
\begin{eqnarray}\label{c1}
\fl\thetaf{l/q+a_A}{b_A}{q\sum_i \Xi_i - \sum_i\xi_i + q}{q\tau}\nonumber\\
=\thetaf{l/q+a_A}{b_A}{q\sum_i \Xi_i-\sum_i\xi_i}{q\tau} \rme^{2\pi \rmi (l+q a_A)}.
\end{eqnarray}
The Jastrow factor transforms like
\begin{eqnarray}
\fl\prod_{i < j} \left( \frac{  E (\Xi_i - \Xi_j + \delta_{ik} - \delta_{jk}, \tau)}{ E (\Xi_i - \Xi_j, \tau)}  \right)^{q} \nonumber \\
= \frac{\prod_{i (< k)} E (\Xi_i - \Xi_k - 1, \tau)^q \prod_{j (> k)} E (\Xi_k - \Xi_j + 1, \tau)^q}{\prod_{i (< k)} E (\Xi_i - \Xi_k, \tau) ^q \prod_{j (> k)} E (\Xi_k - \Xi_j, \tau)^q } \nonumber \\
= (-1)^{\sum_{i(<k)} q + \sum_{j(>k)} q}
= (-1)^{q(N/q - 1)}
= (-1)^{N-q},
\end{eqnarray}
where we have used the property $E(\xi \pm 1, \tau) = - E(\xi, \tau) $, and
\begin{eqnarray}
\prod_{i=1}^M\prod_{j=1}^N \left(\frac{\tilde{E}(\Xi_i + \delta_{ik} - \xi_j, \tau)}{\tilde{E}(\Xi_i - \xi_j, \tau)}\right)^{-1}=(-1)^N.
\end{eqnarray}
The overall transformation is hence
\begin{eqnarray}
\fl\psi_{l/q+a_A,b_A}(\xi_1,\ldots,\xi_N,\Xi_1,\ldots,\Xi_{k-1},\Xi_k+1,\Xi_{k+1},\ldots,\Xi_M)\nonumber\\
=\psi_{l/q+a_A,b_A}(\xi_1,\ldots,\xi_N,\Xi_1,\ldots,\Xi_M)\rme^{2\pi\rmi l}(-1)^{q(2a_A-1)}.
\end{eqnarray}
This shows that for $q$ even, the state is periodic, whereas for $q$ odd it can be periodic ($a_A=1/2$) or antiperiodic ($a_A=0$). The phase factor obtained when taking a particle around the torus in the $\omega_1$ direction measures the flux through the hole of the torus. We thus observe that increasing $l$ by one, increases the flux through the hole by one.

\subsection{$\Xi_k\to \Xi_k+\tau$}

In this case, $q \sum_i \Xi_i \rightarrow q \sum_i \Xi_i + q\tau$. We shall need the properties
\begin{eqnarray}
\thetaf{a}{b}{\xi + p\tau}{\tau} = \rme^{-\rmi\pi\tau p^2 - 2\pi\rmi p(\xi + b) } \thetaf{a+p}{b}{\xi}{\tau}, \quad p \in \mathbb{R},\\
\thetaf{a+n}{b}{\xi}{\tau} = \thetaf{a}{b}{\xi}{\tau}, \quad n \in \mathbb{Z}, \label{thetaprop1}\\
E(\xi \pm \tau, \tau) = \rme^{-\rmi\pi\tau \mp \rmi\pi \mp 2\pi\rmi\xi} E(\xi, \tau).\label{Etau}
\end{eqnarray}
Now
\begin{eqnarray}\label{c2}
\fl\thetaf{l/q+a_A}{b_A}{q\sum_i \Xi_i-\sum_i\xi_i + q\tau}{q\tau} \nonumber\\
=\thetaf{l/q+a_A}{b_A}{q\sum_i \Xi_i-\sum_i\xi_i}{q\tau} \rme^{-\rmi\pi q \tau - 2\pi \rmi \left(q\sum_i \Xi_i-\sum_i\xi_i + b_A\right) }
\end{eqnarray}
and
\begin{eqnarray}
\fl\prod_{i < j} \left( \frac{  E (\Xi_i - \Xi_j+\tau(\delta_{ik}-\delta_{jk}), \tau)}{ E (\Xi_i - \Xi_j, \tau)}  \right)^q \nonumber\\
=\frac{\prod_{i (< k)} E (\Xi_i - \Xi_k - \tau, \tau) ^q \prod_{j (> k)} E (\Xi_k - \Xi_j + \tau, \tau)^q}{\prod_{i (< k)} E (\Xi_i - \Xi_k, \tau)^q \prod_{j (> k)} E (\Xi_k - \Xi_j, \tau)^q} \nonumber \\
= \rme^{\sum_{i (<k)} (-\rmi\pi\tau + \rmi\pi + 2\pi\rmi (\Xi_i-\Xi_k))q} \rme^{\sum_{j (>k)} (-\rmi\pi\tau - \rmi\pi - 2\pi\rmi (\Xi_k-\Xi_j) )q} \nonumber\\
= \rme^{-\rmi\pi \tau q (N/q - 1)} \rme^{\rmi\pi q (N/q - 1) } \rme^{2\pi\rmi q\sum_i\Xi_i - 2\pi\rmi q\Xi_k} \rme^{-2\pi\rmi q(N/q - 1)\Xi_k} \nonumber\\
= \rme^{-\rmi \pi \tau (N - q)} \rme^{\rmi\pi (N - q) } \rme^{2\pi\rmi q\sum_i\Xi_i} \rme^{-2\pi\rmi N\Xi_k}
\end{eqnarray}
and
\begin{eqnarray}
\fl\prod_{i=1}^M\prod_{j=1}^N \left( \frac{ \tilde{E} (\Xi_i + \tau\delta_{ik} - \xi_j, \tau)}{ \tilde{E} (\Xi_i - \xi_j, \tau)}  \right)^{-1} \nonumber\\
=\prod_{j=1}^N (\rme^{-\rmi\pi\tau - \rmi\pi - 2\pi\rmi(\Xi_k-\xi_j)})^{-1}
=\rme^{\rmi\pi\tau N + \rmi\pi N + 2\pi\rmi N\Xi_k-2\pi\rmi\sum_{j=1}^N\xi_j}
\end{eqnarray}
Combining the factors, we get
\begin{eqnarray}
\fl\psi_{l/q+a_A,b_A}(\xi_1,\ldots,\xi_N,\Xi_1,\ldots,\Xi_{k-1},\Xi_k+\tau,\Xi_{k+1},\ldots,\Xi_M)\nonumber\\
=\psi_{l/q+a_A,b_A}(\xi_1,\ldots,\xi_N,\Xi_1,\ldots,\Xi_M)
\rme^{ - 2\pi \rmi b_A-\rmi\pi q}.
\end{eqnarray}
The state is hence periodic for $q$ even, and it is periodic ($b_A=1/2$) or antiperiodic ($b_A=0$) for $q$ odd. We also observe that the flux through the tube of the torus is independent of $l$.

\subsection{Twisted boundary conditions}

From the above derivations it is clear that the states can be modified to have twisted boundary conditions with twist angles $\phi_1$ and $\phi_2$ by modifying the centre-of-mass factor to
\begin{eqnarray}
\thetaf{\phi_1/(2\pi q)+l/q+a_A}{-\phi_2/(2\pi)+b_A}{q\sum_i \Xi_i-\sum_i\xi_i}{q\tau}
\end{eqnarray}
without modifying other parts of the wave functions. In this case
\begin{eqnarray}
\fl\psi_{\phi_1/(2\pi q)+l/q+a_A,-\phi_2/(2\pi)+b_A}(\xi_1,\ldots,\xi_N, \Xi_1,\ldots,\Xi_{k-1},\Xi_k+1,\Xi_{k+1},\ldots,\Xi_M)\nonumber\\
\fl=\psi_{\phi_1/(2\pi q)+l/q+a_A,-\phi_2/(2\pi)+b_A}(\xi_1,\ldots,\xi_N, \Xi_1,\ldots,\Xi_M)\rme^{\rmi \phi_1}\rme^{2\pi\rmi l}(-1)^{q(2a_A-1)}
\end{eqnarray}
and
\begin{eqnarray}
\fl\psi_{\phi_1/(2\pi q)+l/q+a_A,-\phi_2/(2\pi)+b_A}(\xi_1,\ldots,\xi_N, \Xi_1,\ldots,\Xi_{k-1},\Xi_k+\tau,\Xi_{k+1},\ldots,\Xi_M)\nonumber\\
=\psi_{\phi_1/(2\pi q)+l/q+a_A,-\phi_2/(2\pi)+b_A}(\xi_1,\ldots,\xi_N, \Xi_1,\ldots,\Xi_M)
\rme^{\rmi \phi_2}\rme^{ - 2\pi \rmi b_A-\rmi\pi q}.
\end{eqnarray}
A flux unit can be added through the hole or the tube of the torus by increasing $\phi_1$ or $\phi_2$ by $2\pi$.

\section{Transformation properties of the states}\label{sec:trans}

Our next aim is to determine how the states transform under rotations and translations. The transformations $\mathcal{O}$ that we shall consider below can all be expressed in terms of a rearrangement operation $d^{-1}(j)$, where $d$ is a bijective map from $\{1,2,\ldots,N\}$ to $\{1,2,\ldots,N\}$. After the transformation, the number of particles on the site at position $\xi_j$ is $n_{d^{-1}(j)}$. We can, however, equally well state that the number of particles at site $\xi_{d(j)}$ is $n_j$. The transformed state is hence given by
\begin{eqnarray}
\mathcal{O}|\psi_{l/q+a_A,b_A}\rangle =\sum_{n_1,\ldots,n_N}\psi_{l/q+a_A,b_A}(\xi_{d(j)},n_j)|n_1,\ldots,n_N\rangle.
\end{eqnarray}

For the case of rotations, we shall assume that $\tau=\rmi$ and that the lattice is invariant under a $90^\circ$ rotation around the origin, i.e.\ that the sets $\{\rmi\xi_1,\rmi\xi_2,\ldots,\rmi\xi_N\}$ and $\{\xi_1,\xi_2,\ldots,\xi_N\}$ are the same. For the case of translations, we shall assume that the lattice is an $L_x\times L_y$ square lattice with lattice constant $c$. In this case $\omega_1=cL_x$, $\omega_2=\rmi cL_y$, and $\tau=\rmi L_y/L_x$. We shall also use the auxiliary indices $x_j \in \{0,1\ldots L_x-1 \}$ and $y_j \in \{0,1\ldots L_y-1 \}$ that are defined such that the index $j = x_j+y_jL_x+1$. In this notation, the position $\xi_j$ of the $j$'th lattice site is $\xi_j=[x_j - (L_x-1)/2 + i(y_j - (L_y-1)/2 )]/L_x$ if we choose the origin such that $\sum_j\xi_j=0$.

\subsection{Rotation by $90^\circ$}
\label{subsec:rotation}
A rotation by $90^\circ$ is described by the transformation $\xi_{d(j)} = \xi_j/\tau$. Note also that $\tau = \rmi = -1/\tau$. Let us first compute
\begin{eqnarray}
\fl\thetaf{a}{b}{\frac{\xi}{\tau}}{\frac{-q}{\tau}} = \sum_{n \in \mathbb{Z}} \exp \left[ \frac{-\rmi\pi q}{\tau} (n+a)^2 + 2\pi\rmi (n+a)\left( \frac{\xi}{\tau} + b \right) \right] \nonumber \\
= \sum_{k \in \mathbb{Z}} \int_{-\infty}^{\infty}  \exp \left[ 2\pi\rmi kx - \frac{\rmi\pi q}{\tau} (x+a)^2 + 2\pi\rmi (x+a)\left( \frac{\xi}{\tau} + b \right) \right] \mathrm{d}x \nonumber \\
= \sum_{k \in \mathbb{Z}} \sqrt{\frac{\tau}{\rmi q}} \exp \left[ -2\pi\rmi ak + \frac{\rmi\pi \tau}{q} \left( \frac{\xi}{\tau} + b + k \right)^2 \right]. \label{eqtheta-rot}
\end{eqnarray}
Now write $k = qn + m$, where $m \in \{0,1,\ldots q-1 \} $. Equation \eref{eqtheta-rot} becomes
\begin{eqnarray}
\fl\thetaf{a}{b}{\frac{\xi}{\tau}}{\frac{-q}{\tau}} = \sum_{m=0}^{q-1} \sum_{n \in \mathbb{Z}} \sqrt{\frac{\tau}{\rmi q}} \exp \Biggl[ -2\pi\rmi a(qn + m) + \rmi\pi\tau q \left( n + \frac{m+b}{q} \right)^2 + \frac{\rmi\pi \xi^2}{q\tau} \nonumber \\
+ 2\pi\rmi\xi \left(n + \frac{b+m}{q} \right) \Biggr] \nonumber\\
\fl=\sqrt{\frac{\tau}{\rmi q}} \sum_{m=0}^{q-1} \sum_{n \in \mathbb{Z}}\exp \left[\rmi\pi\tau q \left( n + \frac{m+b}{q} \right)^2 + \frac{\rmi\pi \xi^2}{q\tau} + 2\pi\rmi\left(n + \frac{b+m}{q} \right)(\xi - qa) + 2\pi\rmi ab \right]\nonumber\\
=\sqrt{\frac{\tau}{\rmi q}} \rme^{\frac{\rmi\pi \xi^2}{q\tau} + 2\pi\rmi ab} \sum_{m=0}^{q-1} \thetaf{\frac{m+b}{q}}{-qa}{\xi}{q\tau}.  \label{rotatetheta}
\end{eqnarray}
A similar computation gives
\begin{eqnarray}
E \left( \frac{\xi_i-\xi_j}{\tau} , \frac{-1}{\tau} \right)=  \tau^{-1}\rme^{\rmi\pi(\xi_i-\xi_j)^2/\tau} E(\xi_i-\xi_j, \tau),
\end{eqnarray}
and hence
\begin{eqnarray}
\fl\delta_n\prod_{i<j} E \left( \frac{\xi_i-\xi_j}{\tau} , \frac{-1}{\tau} \right)^{qn_in_j - n_i -n_j } = \delta_n \prod_{i<j} \left[ \tau^{-1}\rme^\mathrm{\rmi\pi(\xi_i-\xi_j)^2/\tau} E(\xi_i-\xi_j, \tau) \right]^{qn_in_j - n_i -n_j }\nonumber\\
= \delta_n \tau^{\frac{N^2}{2q} + \frac{N}{2} - \frac{N}{q}} \rme^{-\frac{\mathrm{i}\pi q}{\tau}\left(\sum_in_i\xi_i\right)^2}\prod_{i<j} E(\xi_i-\xi_j, \tau)^{qn_in_j - n_i -n_j },
\end{eqnarray}
where we have used the properties $\sum_i\xi_i = 0 $ and $\sum_i\xi_i^2 = 0$, which follow from the assumption that the lattice is invariant under a $90^\circ$ rotation.

Combining the factors and utilizing $\tau=\rmi$ and \eref{thetaprop2}, we conclude that
\begin{eqnarray}\label{resrot}
\fl\psi_{l/q+a_A,b_A}(\xi_{d(j)}, n_j) = \rmi^{\frac{N^2}{2q} + \frac{N}{2} - \frac{N}{q}} \rme^{2\pi\rmi a_A b_A} \frac{1}{\sqrt{q}} \sum_{m=0}^{q-1} \rme^{-2\pi\rmi lm/q }\psi_{\frac{m+b_A}{q},-qa_A}(\xi_j, n_j).
\end{eqnarray}
In particular, utilizing \eref{thetaprop2} and \eref{thetaprop1}, we obtain
\begin{eqnarray}
\fl\psi_{\frac{l}{q},0}(\xi_{d(j)}, n_j) = \rmi^{\frac{N^2}{2q} + \frac{N}{2} - \frac{N}{q}} \sum_{m=0}^{q-1} \mathcal{R}_{lm} \, \psi_{\frac{m}{q},0}(\xi_j, n_j),\\
\fl\psi_{\frac{l}{q},\frac{1}{2}}(\xi_{d(j)}, n_j) = \rmi^{\frac{N^2}{2q} + \frac{N}{2} - \frac{N}{q}} \rme^{-\pi\rmi l(q-1)/q} \sum_{m=0}^{q-1} \mathcal{R}_{lm} \,\psi_{\frac{m}{q}+\frac{1}{2},0}(\xi_j, n_j),\\
\fl\psi_{\frac{l}{q}+\frac{1}{2},0}(\xi_{d(j)}, n_j) = \rmi^{\frac{N^2}{2q} + \frac{N}{2} - \frac{N}{q}} \sum_{m=0}^{q-1} \mathcal{R}_{lm} \, \rme^{-\pi\rmi m(q+1)/q}\psi_{\frac{m}{q},\frac{1}{2}}(\xi_j, n_j),\\
\fl\psi_{\frac{l}{q}+\frac{1}{2},\frac{1}{2}}(\xi_{d(j)}, n_j) = \rmi^{\frac{N^2}{2q} + \frac{N}{2} - \frac{N}{q}} \rme^{-\pi\rmi l(q-1)/q } \sum_{m=0}^{q-1} \mathcal{R}_{lm}\,
\rme^{-\pi\rmi (m(q+1)/q+q/2) } \psi_{\frac{m}{q}+\frac{1}{2},\frac{1}{2}}(\xi_j, n_j),
\end{eqnarray}
where it is assumed in the last three equations that $q$ is odd and $\mathcal{R}_{lm}=\rme^{-2\pi\rmi lm/q }/\sqrt{q}$ are the elements of the unitary matrix
\begin{eqnarray}\label{Rmatrix}
\mathcal{R}\equiv\frac{1}{\sqrt{q}}
\left[\begin{array}{cccc}
1		& 1						   & \ldots  & 1 \\
1		& \rme^{-2\pi\rmi/q}	   & \ldots	 & \rme^{-2\pi\rmi(q-1)/q} \\
\vdots  & \vdots      			   & \ddots	 & \vdots \\
1		& \rme^{-2\pi\rmi(q-1)/q}  & \ldots	 & \rme^{-2\pi\rmi(q-1)^2/q}
\end{array}\right].
\end{eqnarray}
We observe that states with periodic (antiperiodic) boundary conditions in both directions transform into states with periodic (antiperiodic) boundary conditions in both directions, while states with periodic boundary conditions in one direction and antiperiodic in the other transform into states with antiperiodic boundary conditions in the one direction and periodic in the other.

\subsection{Translation by one lattice site in the $x$ direction}

Translation by one lattice site in the $x$ direction is described by the rearrangement transformation
\begin{eqnarray}
d^{-1}(j) = \left\{ \begin{array}{cc}
j + 1 & \textrm{for } x_j > 0  \\
j - (L_x - 1) & \textrm{for } x_j = 0
\end{array}\right..
\end{eqnarray}
The corresponding change in lattice positions is
\begin{eqnarray}
\xi_{d(j)} =
\left\{\begin{array}{cl}
\xi_j - 1/L_x & \textrm{for } x_j > 0 \\
\xi_j + (L_x - 1)/L_x & \textrm{for } x_j = 0
\end{array}\right..
\end{eqnarray}
This leads to
\begin{eqnarray}
\xi_{d(i)} - \xi_{d(j) } = \left\{ \begin{array}{cl}
\xi_i - \xi_j & \textrm{ for } \ x_i, x_j > 0 \ \textrm{ or } \ x_i, x_j = 0 \\
\xi_i - \xi_j + 1  & \textrm{ for }\  x_i  = 0, x_j > 0 \\
\xi_i - \xi_j - 1  & \textrm{ for }\ x_i  > 0, x_j = 0 \\
\end{array}\right.
\end{eqnarray}
and
\begin{eqnarray}
\sum_i\xi_{d(i) }(qn_i-1) = \sum_i\xi_{i }(qn_i-1) - L_y + qn_L,
\end{eqnarray}
where $n_L \equiv \sum_{\{i|x_i = 0\}} n_i $. Utilizing \eref{thetaprop2} and $E(\xi \pm 1, \tau) = - E(\xi, \tau)$, we obtain
\begin{eqnarray}
\fl\thetaf{a}{b}{\sum_i\xi_{d(i) }(qn_i-1)}{q\tau}
=\thetaf{a}{b}{\sum_i\xi_{i}(qn_i-1)}{q\tau} \rme^{2\pi\rmi a(-L_y + q n_L)}
\end{eqnarray}
and
\begin{eqnarray}
\fl\delta_n\prod_{i < j} E (\xi_{d(i) } - \xi_{d(j) }, \tau) ^{q n_in_j-n_i-n_j}\nonumber\\
=\delta_n\prod_{i < j} E (\xi_{i } - \xi_{j }, \tau) ^{q n_in_j-n_i-n_j}
(-1)^{q n_L (N/q-n_L) - n_L(N-L_y) - L_y(N/q-n_L)}\nonumber \\
=\delta_n\prod_{i < j} E (\xi_{i } - \xi_{j }, \tau) ^{q n_in_j-n_i-n_j}  (-1)^{-qn_L - L_yN/q}.
\end{eqnarray}
Collecting the factors, we get
\begin{eqnarray}
\fl\psi_{l/q+a_A,b_A}(\xi_{d(j)},n_j)=\psi_{l/q+a_A,b_A}(\xi_j,n_j)
\rme^{-2\pi\rmi l L_y/q -2\pi\rmi a_A L_y + 2\pi\rmi a_A q n_L}(-1)^{qn_L+L_yN/q}.
\end{eqnarray}
In particular,
\begin{eqnarray}
\psi_{l/q,0}(\xi_{d(j)},n_j)=\psi_{l/q,0}(\xi_j,n_j)
\,\rme^{-2\pi\rmi l L_y/q}(-1)^{qn_L+L_yN/q},\\
\psi_{l/q,1/2}(\xi_{d(j)},n_j)=\psi_{l/q,1/2}(\xi_j,n_j)
\,\rme^{-2\pi\rmi l L_y/q}(-1)^{qn_L+L_yN/q},\\
\psi_{l/q+1/2,0}(\xi_{d(j)},n_j)=\psi_{l/q+1/2,0}(\xi_j,n_j)
\,\rme^{-2\pi\rmi l L_y/q}(-1)^{L_y(N/q-1)},\\
\psi_{l/q+1/2,1/2}(\xi_{d(j)},n_j)=\psi_{l/q+1/2,1/2}(\xi_j,n_j)
\,\rme^{-2\pi\rmi l L_y/q}(-1)^{L_y(N/q-1)}.
\end{eqnarray}
Note that $(-1)^{qn_L}=1$ for $q$ even. We thus observe that the states $|\psi_{l/q+a_A,b_A}\rangle$ are eigenstates of the translation operator, whenever we choose periodic boundary conditions in the $x$ direction.

\subsection{Translation by one lattice site in the $y$ direction}

The calculation for this case is similar to the previous one. Now we have
\begin{eqnarray}
d^{-1}(j) = \left\{ \begin{array}{cl}
j + L_x & \textrm{for } y_j > 0  \\
j - (L_y - 1)L_x & \textrm{for } y_j = 0
\end{array}\right.
\end{eqnarray}
and
\begin{eqnarray}
\xi_{d(j)}=
\left\{
\begin{array}{cl}
\xi_j - \rmi/L_x & \textrm{for } y_j > 0 \\
\xi_j + \rmi(L_y - 1)/L_x & \textrm{for } y_j = 0
\end{array}\right.,
\end{eqnarray}
leading to
\begin{eqnarray}
\xi_{d(i)} - \xi_{d(j) } = \left\{ \begin{array}{cl}
\xi_i - \xi_j & \textrm{ for } \ y_i, y_j > 0 \ \textrm{ or } \ y_i, y_j = 0\\
\xi_i - \xi_j + \tau  & \textrm{ for }\  y_i = 0, y_j > 0\\
\xi_i - \xi_j - \tau  & \textrm{ for }\  y_i > 0, y_j = 0
\end{array}\right.
\end{eqnarray}
and
\begin{eqnarray}
\sum_i\xi_{d(i) }qn_i = \sum_i\xi_{i }qn_i + q\tau(n_B - L_x/q),
\end{eqnarray}
where $n_B\equiv\sum_{\{j|y_j = 0\}} n_j$. Using the property
\begin{eqnarray}
\thetaf{a}{b}{\xi + p\tau}{\tau} = \rme^{-\rmi\pi\tau p^2 - 2\pi\rmi p(\xi + b)} \thetaf{a + p}{b}{\xi}{\tau} \textrm{ for } p \in \mathbb{R}
\end{eqnarray}
with $p = n_B - L_x / q $, we get
\begin{eqnarray}
\fl\thetaf{l/q+a_A}{b_A}{\sum_i\xi_{d(i) }(qn_i-1)}{q\tau} =
\thetaf{l/q+a_A + n_B - L_x/q}{b_A} {\sum_i\xi_{i}(qn_i-1)}{q\tau}\nonumber\\
\times\rme^{-\rmi\pi q \tau\left(n_B - L_x/q\right)^2 - 2\pi \rmi\left(n_B - L_x/q\right)(\sum_i\xi_i(qn_i-1) + b_A)}.
\end{eqnarray}
Also, from \eref{Etau}, we get
\begin{eqnarray}
\fl \delta_n \prod_{i<j} E(\xi_{d(i)} - \xi_{d(j) },\tau)^{qn_in_j-n_i-n_j}
=\delta_n \, \rme^{-\rmi\pi\sum_{\{i|y_i=0\}}\sum_{\{j|y_j>0\}}(\tau+1+ 2(\xi_i-\xi_j))(qn_in_j-n_i-n_j)} \nonumber\\
\times\prod_{i<j} E(\xi_i - \xi_j,\tau)^{qn_in_j-n_i-n_j}
\end{eqnarray}
After some computations, this gives the transformation
\begin{eqnarray}
\fl\psi_{l/q+a_A,b_A}(\xi_{d(j)}, n_j) = \psi_{(l-L_x)/q+a_A,b_A} (\xi_j, n_j) \; \rme^{ 2\pi\rmi b_A \left( L_x/q - n_B \right)} \, (-1)^{qn_B + NL_x/q}.
\end{eqnarray}
In particular,
\begin{eqnarray}
\psi_{l/q+a_A,0}(\xi_{d(j)}, n_j) = \psi_{(l-L_x)/q+a_A,0}(\xi_j, n_j) \, (-1)^{qn_B + NL_x/q},\\
\psi_{l/q+a_A,1/2}(\xi_{d(j)}, n_j) = \psi_{(l-L_x)/q+a_A,1/2} (\xi_j, n_j) \, (-1)^{(N+1)L_x/q},
\end{eqnarray}
where $q$ odd has been assumed in the last equation. The states with periodic boundary conditions in the $y$ direction and with $L_x/q$ integer are hence eigenstates of the transformation.

\section{Orthonormality and the $S$-matrix}

\begin{figure}
\begin{indented}\item[]
\includegraphics[width=0.41\columnwidth,bb=101 270 477 561]{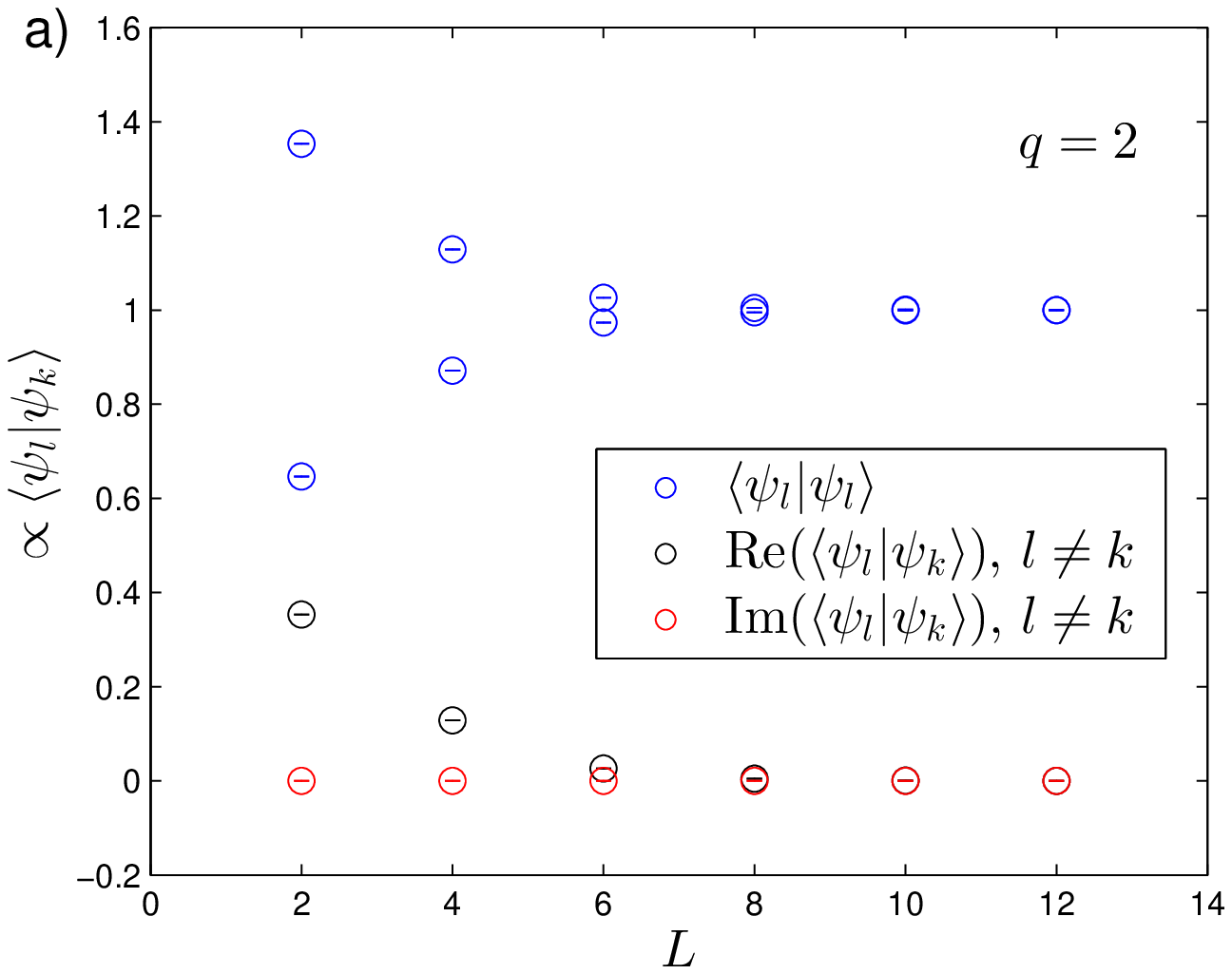}
\includegraphics[width=0.41\columnwidth,bb=101 270 477 561]{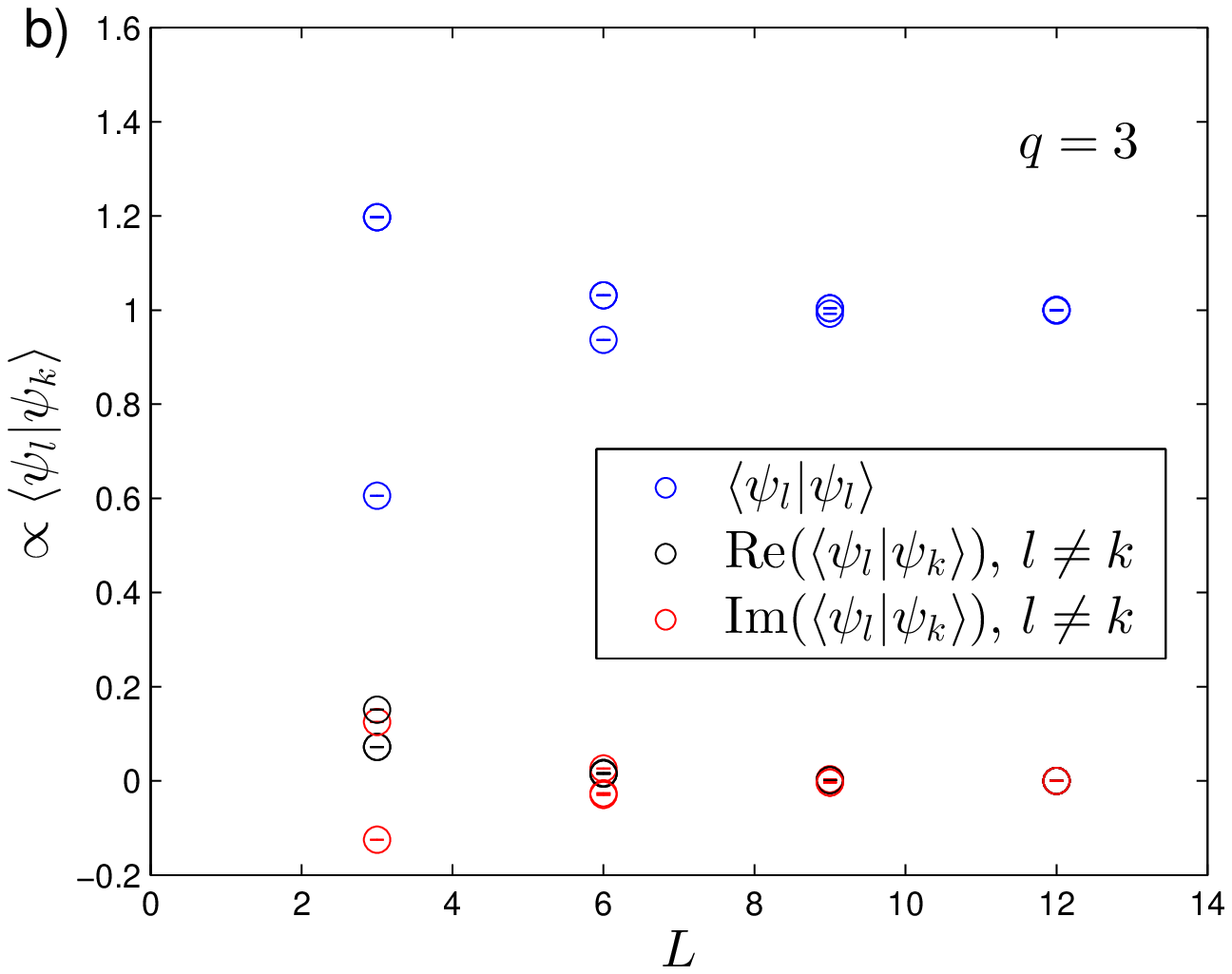}
\caption{Overlap $\langle\psi_l|\psi_k\rangle$ of the states on an $L\times L$ lattice for (a) $q=2$ and (b) $q=3$ computed with Monte Carlo simulations, where $|\psi_l\rangle\equiv|\psi_{l/q+a_A,b_A}\rangle$ with $a_A=b_A=0$ for $q$ even and $a_A=b_A=1/2$ for $q$ odd. The overlaps in the figure are scaled such that $\sum_{l=0}^{q-1}\langle\psi_l|\psi_l\rangle=q$. It is seen that the states $|\psi_l\rangle$ become orthogonal and have the same norm for large $L$. They can hence be made orthonormal by multiplying all the $q$ states with the same factor. A similar behaviour is found for the other possible choices of $a_A$ and $b_A$ for $q=3$.} \label{fig:overlap}
\end{indented}
\end{figure}

In particular cases, it is possible to utilize the transformation properties derived in the previous section to analytically find linear combinations of the states that are orthogonal to each other, but it is not guaranteed in general that the states obtained from CFT are orthogonal. In figure \ref{fig:overlap}, we study the overlap between the states $|\psi_{l/q+a_A,b_A}\rangle$ on an $L\times L$ square lattice with periodic boundary conditions numerically for $q=2$ and $q=3$. We do this by rewriting the overlaps into an expectation value
\begin{eqnarray}
\fl\langle\psi_{l/q+a_A,b_A}|\psi_{k/q+a_A,b_A}\rangle=\sum_{n_1,\ldots,n_N} \frac{\overline{F_{l,A}^\mathrm{cm}(n_1,\ldots,n_N)}F_{k,A}^\mathrm{cm}(n_1,\ldots,n_N)} {\sum_p |F_{p,A}^\mathrm{cm}(n_1,\ldots,n_N)|^2} P_A(n_1,\ldots,n_N)
\end{eqnarray}
where
\begin{eqnarray}
F_{l,A}^\mathrm{cm}(n_1,\ldots,n_N)\equiv\thetaf{l/q+a_A}{b_A}{\sum_{i=1}^N \xi_i (qn_i - 1)}{q\tau}
\end{eqnarray}
is the centre-of-mass factor and
\begin{eqnarray}
\fl P_A(n_1,\ldots,n_N)\equiv \delta_{n} \sum_p |F_{p,A}^\mathrm{cm}(n_1,\ldots,n_N)|^2
\left|\prod_{i < j} E(\xi_i - \xi_j, \tau)^{qn_in_j -n_i- n_j }\right|^2
\end{eqnarray}
can be interpreted as a probability distribution for the configurations $n_1,\ldots,n_N$. This allows us to compute $\langle\psi_{l/q+a_A,b_A}|\psi_{k/q+a_A,b_A}\rangle$ up to a constant that does not depend on $l$ and $k$ using a Metropolis Monte Carlo algorithm. The figure shows that the states become orthogonal for large lattices. In addition, all the states have the same norm, so that they can be made orthonormal by multiplying by a common factor. Since the states $|\psi_{l/q+a_A,b_A}\rangle$ have a well-defined flux through the hole and the tube of the torus, it follows that the $S$-matrix for large lattices can be obtained directly from the transformation properties of the states under a $90^\circ$ rotation \cite{quasiparticle}. This gives us that the $S$-matrix is the matrix $\mathcal{R}$ in \eref{Rmatrix} (up to unimportant permutations of the rows and columns caused by the factors $\rme^{-\pi\rmi l(q-1)/q}$ and $\rme^{-\pi\rmi m(q+1)/q}$). This means that the phase factor acquired by the wave function when an anyon of type $i$ is moved adiabatically in a closed path around an anyon of type $j$ is $\exp(-2\pi\rmi p_ip_j/q)$, where $p_k\in\{0,1,\ldots,q-1\}$ and $p_k/q$ is the charge of an anyon of type $k$. We thus observe that the statistics of the anyons are as expected for the Laughlin states with filling factor $1/q$.

\section{Connection to the Laughlin states on the torus}

We shall now compare the lattice states obtained from CFT to the continuum Laughlin states on a torus \cite{haldane85a,haldane85b,read96} in the Landau gauge (here we use the expression in \cite{read96})
\begin{eqnarray}\label{stateLT}
\fl|\psi_{l,\mathrm{LT}}\rangle\propto\int d^2Z_1\cdots\int d^2Z_M \;\;
\thetaf{l/q + (M-1)/2 }{-(M-1)q/2 }{\sum_i \frac{qZ_i}{\omega_1}}{q\tau}
\prod_{i < j} E\left(\frac{Z_i - Z_j}{\omega_1}, \tau\right)^{q} \nonumber\\
\times\prod_{i=1}^M\rme^{-(1/2)(\mathrm{Im}(Z_i))^2 } \;
|Z_1,\ldots,Z_M\rangle, \quad l\in\{0,1,\ldots,q-1\}.
\end{eqnarray}
In this section, we shall often use the unscaled coordinates $z_i$ and $Z_i$ instead of the scaled coordinates $\xi_i=z_i/\omega_1$ and $\Xi_i=Z_i/\omega_1$. As in previous sections, lower case letters refer to the positions of the lattice sites, while upper case letters refer only to the positions of the occupied lattice sites. In the Landau gauge, the coefficients of the wave functions in \eref{stateLT} do not change when $Z_i\to Z_i+w_1$, and it changes by a $Z_i$-dependent phase factor when $Z_i\to Z_i+w_2$ (see e.g.\ \cite{chung07} for details).

Let us consider the case of an $L_x\times L_y$ square lattice with $\sum_i\xi_i=0$ and lattice constant $\sqrt{2\pi}$ and choose $\omega_1=\sqrt{2\pi}L_x$ and $\omega_2=\rmi\sqrt{2\pi}L_y$. In section 4.5 of \cite{torus}, it has been shown that for this lattice and $i\in\{1,2,\ldots,N\}$, we have
\begin{eqnarray}\label{Esq}
\prod_{j=1}^N \tilde{E}(\xi_{x_i+y_iL_x+1} - \xi_j, \tau)^{-1} \propto
(-1)^{x_iL_y+y_i} \rme^{-\pi(y_i-(L_y-1)/2 )^2 }.
\end{eqnarray}
(The proof in \cite{torus} is for $N$ even, but it can easily be generalized to apply for both even and odd $N$.) Note in particular that $\xi_{x_i+y_iL_x+1}\in\{\xi_1,\xi_2,\ldots,\xi_N\}$, which means that the left and right hand side of \eref{Esq} do not necessarily transform in the same way when the $i$th particle is moved by $w_1$ or $w_2$. Under the assumption that $Z_i\in\{z_1,z_2,\ldots,z_N\}$ for all $i$, we can use \eref{Esq} to rewrite \eref{statep} into
\begin{eqnarray}\label{statepsq}
\fl|\psi_{l/q+a_A,b_A}\rangle\propto\sum_{\{Z_1,\ldots,Z_M\}}
\thetaf{l/q+a_A}{b_A}{\sum_{i=1}^M\frac{qZ_i}{\sqrt{2\pi}L_x}}{q\tau}
\prod_{i < j} E\left(\frac{Z_i - Z_j}{\sqrt{2\pi}L_x}, \tau\right)^q \nonumber\\
\times\prod_{i=1}^M (-1)^{X_iL_y + Y_i} \; \prod_{i=1}^M \rme^{-(1/2)(\mathrm{Im}(Z_i))^2 } \; a_{Z_1}^\dag a_{Z_2}^\dag \cdots a_{Z_M}^\dag|0\rangle,
\end{eqnarray}
where $X_i$ and $Y_i$ are defined such that $Z_i = \sqrt{2\pi}[X_i - (L_x-1)/2 + i(Y_i - (L_y-1)/2 )]$. Utilizing \eref{aAbA} and the properties \eref{thetaprop2} and \eref{thetaprop1} of the theta functions, we observe that the centre-of-mass factors in \eref{stateLT} and \eref{statepsq} are the same if
\begin{eqnarray}\label{choiceA}
A=\left\{\begin{array}{cl}
\mathrm{I} & \textrm{for }q\textrm{ even}\\
\mathrm{I} & \textrm{for }q\ \& \ N \textrm{ odd}\\
\mathrm{IV} & \textrm{for }q\textrm{ odd}\ \& \ N \textrm{ even}
\end{array}\right.
\end{eqnarray}
and $l\to l+q/2\textrm{ mod }q$ if both $q$ and $M$ are even. Note that $N$ can only be odd for $q$ odd, since the number of particles $M=N/q$ must be an integer. The single particle phase factors $\prod_{i=1}^M (-1)^{X_iL_y + Y_i}=\prod_{i=1}^N (-1)^{(x_iL_y + y_i)n_i}$ can be removed with a simple transformation if desired. The states in \eref{statepsq} with the choice \eref{choiceA} are hence practically the Laughlin states, except that the possible positions of the particles are restricted to a lattice. The states with $A$ equal to $\textrm{II}$, $\textrm{III}$, or $\textrm{IV}$, or to $\textrm{I}$, $\textrm{II}$, and $\textrm{III}$ for $q$ odd have boundary conditions that differ by a sign in one or both of the two directions as can be seen from \eref{c1} and \eref{c2}.

\section{Conclusion}

In summary, we have started from a CFT correlator of a product of vertex operators and used it to derive analytical expressions for a family of lattice states on the torus that are closely related to the bosonic and fermionic Laughlin states at filling factor $1/q$. We have computed various properties of the states and also shown how the states are related to the Laughlin states. The same CFT correlator has previously been used to derive Laughlin-like models on the plane and the cylinder \cite{genq}, and the present work provides the corresponding wave functions on the torus. Finding the states on the torus is interesting because it opens up possibilities to use a number of tools to study the topological properties of the states. Here, we have used some of these to find the $S$-matrix, which shows that the braiding properties of the anyons are the same as for the Laughlin states. These results and previous investigations of the states in the cylinder and plane geometry \cite{genq,nielsen,rodriguez} show that the presence of the lattice and the small differences in analytical expressions between the CFT states and the Laughlin states do not change the topological properties of the states. In addition, the states on the torus are helpful for numerical studies of, e.g., the bulk properties of the states.

\ack
The authors would like to thank Germ\'an Sierra for discussions. AD wishes to thank the entire theory group at MPQ for their warm hospitality. This work has been supported by the EU project SIQS and by the Villum Foundation.

\section*{References}

\bibliographystyle{unsrt}
\bibliography{bibfil}

\begin{thebibliography}{10}

\bibitem{wf}
R.~B. Laughlin.
\newblock Anomalous quantum {H}all effect: An incompressible quantum fluid with
  fractionally charged excitations.
\newblock {\em Phys. Rev. Lett.}, 50:1395--1398, May 1983.

\bibitem{hierarchy1}
F.~D.~M. Haldane.
\newblock Fractional quantization of the {H}all effect: A hierarchy of
  incompressible quantum fluid states.
\newblock {\em Phys. Rev. Lett.}, 51:605--608, Aug 1983.

\bibitem{hierarchy2}
B.~I. Halperin.
\newblock Statistics of quasiparticles and the hierarchy of fractional
  quantized {H}all states.
\newblock {\em Phys. Rev. Lett.}, 52:1583--1586, Apr 1984.

\bibitem{haldane85a}
F.~D.~M. Haldane and E.~H. Rezayi.
\newblock Periodic {L}aughlin-{J}astrow wave functions for the fractional
  quantized {H}all effect.
\newblock {\em Phys. Rev. B}, 31:2529--2531, Feb 1985.

\bibitem{haldane85b}
F.~D.~M. Haldane.
\newblock Many-particle translational symmetries of two-dimensional electrons
  at rational {L}andau-level filling.
\newblock {\em Phys. Rev. Lett.}, 55:2095--2098, Nov 1985.

\bibitem{cf}
J.~K. Jain.
\newblock Composite-fermion approach for the fractional quantum {H}all effect.
\newblock {\em Phys. Rev. Lett.}, 63:199--202, Jul 1989.

\bibitem{CFTstates}
G.~Moore and N.~Read.
\newblock Nonabelions in the fractional quantum {H}all effect.
\newblock {\em Nucl. Phys. B}, 360(2):362--396, 1991.

\bibitem{chung07}
S.~B. Chung and M.~Stone.
\newblock Explicit monodromy of {M}oore--{R}ead wavefunctions on a torus.
\newblock {\em Journal of Physics A: Mathematical and Theoretical}, 40:4923,
  2007.

\bibitem{hermanns}
M.~Hermanns, J.~Suorsa, E.~J. Bergholtz, T.~H. Hansson, and A.~Karlhede.
\newblock Quantum {H}all wave functions on the torus.
\newblock {\em Phys. Rev. B}, 77:125321, Mar 2008.

\bibitem{hermanns2}
M.~Hermanns.
\newblock Composite fermion states on the torus.
\newblock {\em Phys. Rev. B}, 87:235128, Jun 2013.

\bibitem{jaksch2003}
D.~Jaksch and P.~Zoller.
\newblock Creation of effective magnetic fields in optical lattices: the
  {H}ofstadter butterfly for cold neutral atoms.
\newblock {\em New Journal of Physics}, 5(1):56, 2003.

\bibitem{PRA.70.041603}
E.~J. Mueller.
\newblock Artificial electromagnetism for neutral atoms: {E}scher staircase and
  {L}aughlin liquids.
\newblock {\em Phys. Rev. A}, 70:041603, Oct 2004.

\bibitem{PRL.94.086803}
A.~S. S\o{}rensen, E.~Demler, and M.~D. Lukin.
\newblock Fractional quantum {H}all states of atoms in optical lattices.
\newblock {\em Phys. Rev. Lett.}, 94:086803, Mar 2005.

\bibitem{hafezi2008}
M.~Hafezi, A.~S. S{\o}rensen, M.~D. Lukin, and E.~Demler.
\newblock Characterization of topological states on a lattice with {C}hern
  number.
\newblock {\em EPL (Europhysics Letters)}, 81(1):10005, 2008.

\bibitem{PhysRevLett.96.180407}
R.~N. Palmer and D.~Jaksch.
\newblock High-field fractional quantum {H}all effect in optical lattices.
\newblock {\em Phys. Rev. Lett.}, 96:180407, May 2006.

\bibitem{PhysRevA.82.043629}
L.~Mazza, M.~Rizzi, M.~Lewenstein, and J.~I. Cirac.
\newblock Emerging bosons with three-body interactions from spin-1 atoms in
  optical lattices.
\newblock {\em Phys. Rev. A}, 82:043629, Oct 2010.

\bibitem{cooper}
N.~R. Cooper and J.~Dalibard.
\newblock Reaching fractional quantum {H}all states with optical flux lattices.
\newblock {\em Phys. Rev. Lett.}, 110:185301, Apr 2013.

\bibitem{yao}
N.~Y. Yao, A.~V. Gorshkov, C.~R. Laumann, A.~M. L\"auchli, J.~Ye, and M.~D.
  Lukin.
\newblock Realizing fractional {C}hern insulators in dipolar spin systems.
\newblock {\em Phys. Rev. Lett.}, 110:185302, Apr 2013.

\bibitem{nc}
A.~E.~B. Nielsen, G.~Sierra, and J.~I. Cirac.
\newblock Local models of fractional quantum {H}all states in lattices and
  physical implementation.
\newblock {\em Nature Communications}, 4:2864, 2013.

\bibitem{sterdyniak}
A.~Sterdyniak, B.~A. Bernevig, N.~R. Cooper, and N.~Regnault.
\newblock Interacting bosons in topological optical flux lattices.
\newblock {\em Phys. Rev. B}, 91:035115, Jan 2015.

\bibitem{neupert}
T.~Neupert, L.~Santos, C.~Chamon, and C.~Mudry.
\newblock Fractional quantum {H}all states at zero magnetic field.
\newblock {\em Phys. Rev. Lett.}, 106:236804, Jun 2011.

\bibitem{sheng}
D.~N. Sheng, Z.-C. Gu, K.~Sun, and L.~Sheng.
\newblock Fractional quantum {H}all effect in the absence of {L}andau levels.
\newblock {\em Nat. Commun.}, 2:389, 2011.

\bibitem{wang}
Y.-F. Wang, Z.-C. Gu, C.-D. Gong, and D.~N. Sheng.
\newblock Fractional quantum {H}all effect of hard-core bosons in topological
  flat bands.
\newblock {\em Phys. Rev. Lett.}, 107:146803, Sep 2011.

\bibitem{regnault}
N.~Regnault and B.~A. Bernevig.
\newblock Fractional {C}hern insulator.
\newblock {\em Phys. Rev. X}, 1:021014, Dec 2011.

\bibitem{schroeter}
D.~F. Schroeter, E.~Kapit, R.~Thomale, and M.~Greiter.
\newblock Spin {H}amiltonian for which the chiral spin liquid is the exact
  ground state.
\newblock {\em Phys. Rev. Lett.}, 99:097202, Aug 2007.

\bibitem{thomale}
R.~Thomale, E.~Kapit, D.~F. Schroeter, and M.~Greiter.
\newblock Parent {H}amiltonian for the chiral spin liquid.
\newblock {\em Phys. Rev. B}, 80:104406, Sep 2009.

\bibitem{kapit}
E.~Kapit and E.~Mueller.
\newblock Exact parent {H}amiltonian for the quantum {H}all states in a
  lattice.
\newblock {\em Phys. Rev. Lett.}, 105:215303, Nov 2010.

\bibitem{prl}
A.~E.~B. Nielsen, J.~I. Cirac, and G.~Sierra.
\newblock Laughlin spin-liquid states on lattices obtained from conformal field
  theory.
\newblock {\em Phys. Rev. Lett.}, 108:257206, Jun 2012.

\bibitem{son}
H.-H. Tu.
\newblock Projected {BCS} states and spin {H}amiltonians for the {SO}($n$)$_1$
  {W}ess-{Z}umino-{W}itten model.
\newblock {\em Phys. Rev. B}, 87:041103, Jan 2013.

\bibitem{greiter}
M.~Greiter, D.~F. Schroeter, and R.~Thomale.
\newblock Parent {H}amiltonian for the non-{A}belian chiral spin liquid.
\newblock {\em Phys. Rev. B}, 89:165125, Apr 2014.

\bibitem{genq}
H.-H. Tu, A.~E.~B. Nielsen, J.~I. Cirac, and G.~Sierra.
\newblock Lattice {L}aughlin states of bosons and fermions at filling fractions
  1/q.
\newblock {\em New Journal of Physics}, 16(3):033025, 2014.

\bibitem{sun}
H.-H. Tu, A.~E.~B. Nielsen, and G.~Sierra.
\newblock Quantum spin models for the {SU}(n)$_1$ {W}ess-{Z}umino-{W}itten
  model.
\newblock {\em Nuclear Physics B}, 886(0):328 -- 363, 2014.

\bibitem{bondesan}
R.~Bondesan and T.~Quella.
\newblock Infinite matrix product states for long-range {SU(N)} spin models.
\newblock {\em Nuclear Physics B}, 886(0):483 -- 523, 2014.

\bibitem{nielsen}
A.~E.~B. Nielsen.
\newblock Anyon braiding in semianalytical fractional quantum {H}all lattice
  models.
\newblock {\em Phys. Rev. B}, 91:041106, Jan 2015.

\bibitem{glasser15}
I.~Glasser, J.~I. Cirac, G.~Sierra, and A.~E.~B. Nielsen.
\newblock Exact parent {H}amiltonians of bosonic and fermionic {M}oore-{R}ead
  states on lattices and local models.
\newblock {\em arXiv:1505.04998}, 2015.

\bibitem{rodriguez}
I.~D. Rodriguez and A.~E.~B. Nielsen.
\newblock Continuum limit of lattice models with {L}aughlin-like ground states
  containing quasiholes.
\newblock {\em arXiv:1506.02683}, 2015.

\bibitem{idmps}
J.~I. Cirac and G.~Sierra.
\newblock Infinite matrix product states, conformal field theory, and the
  {H}aldane-{S}hastry model.
\newblock {\em Phys. Rev. B}, 81:104431, Mar 2010.

\bibitem{su2k}
A.~E.~B. Nielsen, J.~I. Cirac, and G.~Sierra.
\newblock Quantum spin {H}amiltonians for the {SU}(2)$_k$ {WZW} model.
\newblock {\em J. Stat. Mech. Theor. Exp.}, 2011(11):P11014, 2011.

\bibitem{torus}
A.~E.~B. Nielsen and G.~Sierra.
\newblock Bosonic fractional quantum {H}all states on the torus from conformal
  field theory.
\newblock {\em Journal of Statistical Mechanics: Theory and Experiment},
  2014(4):P04007, 2014.

\bibitem{kitaev}
A.~Kitaev and J.~Preskill.
\newblock Topological entanglement entropy.
\newblock {\em Phys. Rev. Lett.}, 96:110404, Mar 2006.

\bibitem{levin}
M.~Levin and X.-G. Wen.
\newblock Detecting topological order in a ground state wave function.
\newblock {\em Phys. Rev. Lett.}, 96:110405, Mar 2006.

\bibitem{jiang}
H.-C. Jiang, Z.~Wang, and L.~Balents.
\newblock Identifying topological order by entanglement entropy.
\newblock {\em Nature Phys.}, 8(12):902--905, 2012.

\bibitem{quasiparticle}
Y.~Zhang, T.~Grover, A.~Turner, M.~Oshikawa, and A.~Vishwanath.
\newblock Quasiparticle statistics and braiding from ground-state entanglement.
\newblock {\em Phys. Rev. B}, 85(23):235151, 2012.

\bibitem{cincio}
L.~Cincio and G.~Vidal.
\newblock Characterizing topological order by studying the ground states on an
  infinite cylinder.
\newblock {\em Phys. Rev. Lett.}, 110:067208, Feb 2013.

\bibitem{zaletel}
M.~P. Zaletel, R.~S.~K. Mong, and F.~Pollmann.
\newblock Topological characterization of fractional quantum {H}all ground
  states from microscopic {H}amiltonians.
\newblock {\em Phys. Rev. Lett.}, 110:236801, Jun 2013.

\bibitem{tu13}
H.-H. Tu, Y.~Zhang, and X.-L. Qi.
\newblock Momentum polarization: {A}n entanglement measure of topological spin
  and chiral central charge.
\newblock {\em Phys. Rev. B}, 88(19):195412, 2013.

\bibitem{orus}
R.~Or\'us, T.-C. Wei, O.~Buerschaper, and A.~Garc\'{i}a-Saez.
\newblock Topological transitions from multipartite entanglement with tensor
  networks: {A} procedure for sharper and faster characterization.
\newblock {\em Phys. Rev. Lett.}, 113:257202, Dec 2014.

\bibitem{CFTbook}
P.~Di Francesco, P.~Mathieu, and D.~S\'en\'echal.
\newblock {\em Conformal Field Theory}.
\newblock Springer-Verlag New York, 1997.

\bibitem{block}
S.~Ribault.
\newblock Conformal field theory on the plane.
\newblock {\em arXiv:1406.4290}, 2014.

\bibitem{cortorus1}
R.~Dijkgraaf, E.~Verlinde, and H.~Verlinde.
\newblock ${C}=1$ conformal field theories on {R}iemann surfaces.
\newblock {\em Communications in Mathematical Physics}, 115(4):649--690, 1988.

\bibitem{cortorus2}
L.~Alvarez-Gaume, J.-B. Bost, G.~Moore, P.~Nelson, and C.~Vafa.
\newblock Bosonization on higher genus {R}iemann surfaces.
\newblock {\em Communications in Mathematical Physics}, 112(3):503--552, 1987.

\bibitem{read96}
N.~Read and E.~Rezayi.
\newblock Quasiholes and fermionic zero modes of paired fractional quantum
  {H}all states: {T}he mechanism for non-{A}belian statistics.
\newblock {\em Phys. Rev. B}, 54:16864--16887, Dec 1996.

\end{thebibliography}

\end{document}